\documentclass{elsarticle}
\usepackage{latexsym}
\usepackage{float}
\usepackage[latin2]{inputenc}
\usepackage{graphicx}
\usepackage{ae}
\usepackage{aecompl}
\usepackage{setspace}
\usepackage{amsmath}
\usepackage{amssymb}
\usepackage{amsthm}
\usepackage[hidelinks]{hyperref}

\makeatletter
\def\ps@pprintTitle{%
	\let\@oddhead\@empty
	\let\@evenhead\@empty
	\def\@oddfoot{\centerline{\thepage}}%
	\let\@evenfoot\@oddfoot}
\makeatother

\begin{document}

\title{On the Border of the Amyloidogenic Sequences: Prefix Analysis of the Parallel Beta Sheets in the PDB\_Amyloid Collection}
	
\author[p]{Kristóf Takács}
\ead{takacs@pitgroup.org}
\author[p,u]{Vince Grolmusz\corref{cor1}}
\ead{grolmusz@pitgroup.org}
\cortext[cor1]{Corresponding author}
\address[p]{PIT Bioinformatics Group, Eötvös University, H-1117 Budapest, Hungary}
\address[u]{Uratim Ltd., H-1118 Budapest, Hungary}

\date{}

\begin{abstract}
	The Protein Data Bank (PDB) today contains more than 153,000 entries with the 3-dimensional structures of biological macromolecules. Using the rich resources of this repository, it is possible identifying subsets with specific, interesting properties for different applications. Our research group prepared an automatically updated list of amyloid- and probably amyloidogenic molecules, the PDB\_Amyloid collection, which is freely available at the address \url{http://pitgroup.org/amyloid}. This resource applies exclusively the geometric properties of the steric structures for identifying amyloids. In the present contribution, we analyze the starting (i.e., prefix) subsequences of the characteristic, parallel beta-sheets of the structures in the PDB\_Amyloid collection, and identify further appearances of these length-5 prefix subsequences in the whole PDB data set. We have identified this way numerous proteins, whose normal or irregular functions involve amyloid formation, structural misfolding, or anti-coagulant properties, simply by containing these prefixes: including the T-cell receptor (TCR), bound with the major histocompatibility complexes MHC-1 and MHC-2; the p53 tumor suppressor protein; a mycobacterial RNA polymerase transcription initialization complex; the human bridging integrator protein BIN-1;  and the tick anti-coagulant peptide TAP.
\end{abstract}

\maketitle

\bigskip
\noindent Running head: On the Borders of the Amyloidogenic Sequences

\medskip
	
\section*{Introduction}

Amyloids are misfolded protein aggregates, which are present in numerous biological organisms as structural building blocks or immunological agents \cite{Gebbink2005,Blanco2012,Iconomidou2008,Falabella2012,Maji2009,Taricska2020,Horvath2019a}. In humans, the amyloid formation is frequently associated with neurodegenerative diseases and abnormal metabolic conditions 
\cite{Alzheimer1907,Prescott2014,Ma2002,Zheng2007}. 

The structural studies of amyloid aggregates were considered to be difficult until recently, since being aggregates, they cannot be crystallized and measured by X-ray diffractometry. With the recent developments of solid-state NMR and cryo-electron microscopy, dozens of amyloid structures were deposited in the Protein Data Bank (PDB) \cite{PDB-base,Stankovic2017} in the past several years. 

With the more than 100 amyloid structures among the PDB's 156 thousand entries, it is now possible to define structural characteristics, which well-describe amyloid structures. One good approach was made by \cite{Stankovic2017}, where the authors, with the application of a combination of textual search and specific geometric conditions, successfully retrieved the known amyloid structures from the PDB.

In a recent work of ours \cite{Takacs2019}, we have defined a geometric set of constraints, by which we selected all the amyloid molecules, found by the method of \cite{Stankovic2017}, plus numerous globular proteins, with partial amyloid-like substructures. We emphasize that we were using geometric constraints for the $\beta$-sheet regions in the coordinate sections of the PDB files, without {\em any} textual search in the annotation section of those files. Since the annotation sections of PDB files are known to be less reliable than the coordinate sections, this technique increases the reliability of our results, and, additionally, helps in devising the proper definition of the amyloid-like structures. The resulting selection of the PDB entries, called the PDB\_Amyloid list, is available as an automatically and regularly updated list of PDB entries, at the site \url{http://pitgroup.org/amyloid/}. Since, on the average, around 30 new PDB entries are deposited every day, the ``automatic update'' feature is clearly necessary for this service. The PDB\_Amyloid list contains more than 640 entries today. 

The geometric constraints, applied in \cite{Takacs2019}, are as follows:

\begin{itemize}
	\item[(i)] First, parallel $\beta$-sheet segments are identified. The parallel segments need to be on separate polypeptide chains, their distance needs to be between 2 and 15 \AA, and the standard deviation of their distance needs to be less than 1.5 \AA.
	
	\item[(ii)] Second, the large curvature parallel segments are excluded;
	
	\item[(iii)] Third, the parallel segments need to cover at least the one-seventh of the length of the whole chain.	
	
\end{itemize}

For a more detailed mathematical description of the constraints above, we refer to the original article \cite{Takacs2019}.

We note that the requirement of considering only segments on separate polypeptide chains efficiently excludes hairpin and $\beta$-barrel structures, and also partial molecular structures, labeled as ``amyloids'', but lacking the repeated, parallel $\beta$-sheets in the PDB-deposited files.

\subsection*{Prefixes}

The more than 640 PDB entries, available at \url{http://pitgroup.org/amyloid/}, yield a rich set of amyloid-related molecular structures. The list of the globular proteins (i.e., not the misfolded, aggregated amyloid structures) have a special feature: These molecules remained soluble, but they have partial sub-structures, satisfying the conditions (i), (ii) and (iii) above. We believe that the sequence-borders of the $\beta$-sheets in these structures have specific roles in the prevention of the transitions to amyloid state: in the globular proteins, these border-regions may prevent or regulate the formation of aggregated amyloid structures from the protein.

In the present contribution, we consider the prefixes of the parallel $\beta$-sheets of the entries of the PDB\_Amyloid list \url{http://pitgroup.org/amyloid/}. These prefixes are the starting subsections,  where the order of the residues is the default N-terminus through C-terminus. 

Here we identify the most frequently found prefixes in the PDB\_Amyloid list, and then we search for them in the {\em complete} Protein Data Bank. We identify numerous interesting hits, which have proven connections to amyloid formation. We stress that the hits, analyzed below, are found by sequence-searches in the whole PDB, without using {\em any} additional structural constraints, where the sequences we searched for were the prefixes, identified in the PDB\_Amyloid list.  

\section*{Methods}

First, the defining parallel $\beta$-sheets, satisfying the conditions (i), (ii) and (iii) in the Introduction, of the PDB\_Amyloid list \url{http://pitgroup.org/amyloid/}, were identified. 

Next, we collected the length-5 prefixes of the form XXYYY, where the first three residues of the parallel $\beta$-sheet are YYY, and the last two residues, preceding the parallel section of the $\beta$-sheet, are XX.

Figure 1 depicts the GLN-LYS-LEU-VAL-PHE (QKLVF) prefix from the amyloid structure of the PDB entry 2MPZ.

\begin{figure}[H]
	\begin{center}
		\includegraphics[width=12cm]{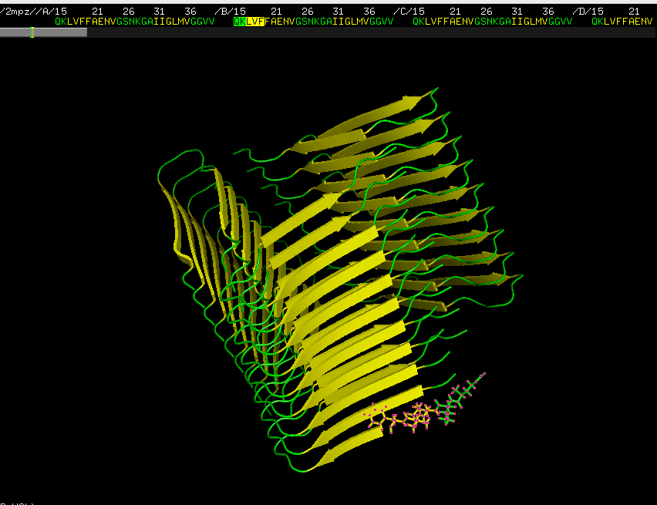}
		\caption{PDB entry 2MPZ, depicted with PyMol. Yellow color denotes $\beta$-sheets. The QKLVF prefix, where QK is green (it is not a part of the parallel $\beta$-sheet) and LVF is yellow (i.e., LVF are the first three residues of the parallel $\beta$-segment), is emphasized at the right bottom of the figure, while its corresponding sequence at the top center. }
	\end{center}
\end{figure}

Next, we have counted the number of appearances of the prefixes and the suffixes in the parallel $\beta$-sheet segments in the PDB\_Amyloid list \url{http://pitgroup.org/amyloid/}. Note that one PDB entry may contain more than one identical prefixes (like in the case of PDB entry 2MPZ, shown in Figure 1). Therefore, the prefix and suffix counts contain multiplicities of two types: (i) multiple appearances in the very same PDB entry, or (ii) multiple appearances in different -- and possibly homologous -- PDB entries. Instead of introducing an unnecessarily complex homogeneity-corrected counting method for the prefix- and suffix appearances in the PDB\_Amyloid list, we just count their raw, uncorrected number of appearances. Since the inclusion or exclusion of the protein structures in the PDB mostly relate to the interest of researchers depositing the structures, and do not carry a statistical or biological meaning. Moreover, we do not count the appearances in the whole PDB, just in the amyloid-like sublist of PDB\_Amyloid. These counts (either corrected or uncorrected) can only be used informally, and do not show the frequency of these subsequences in the protein structures in Nature.

\section*{Discussion and results}

In what follows, we consider the PDB\_Amyloid list and note if the prefix appears in structures, described by the application of NMR spectroscopy (both solid and liquid phase), or by X-ray diffractometry. 

\subsection*{The QKLVF Prefix} 
The QKLVF prefix (i.e., GLN, LYS, LEU, VAL, PHE) appears 77 times in the NMR-identified members of the PDB\_Amyloid list, in the following PDB-structures: 2LMN, 2LMO, 2LMP, 2LMQ, 2LNQ, 2MPZ. The 2MPZ structure is depicted in Figure 1. We are interested in the appearances of the QKLVF subsequence in the whole PDB, and we intend to identify the protein structures, which have the proven potential to turn to amyloids. 

Among the numerous $\beta$-amyloid hits, which are not reviewed here, several interesting appearances of the QKLVF subsequence in globular proteins are in the T-cell receptors (TCR), bound with the major histocompatibility complexes MHC-1 and MHC-2, in the PDB entries 4P5T, 4OZF, 3QIU, 3QIW, 1BD2, 2IAN, 2IAM, 2IAL, 4WW1, 4WW2, 5XOT.  Very interestingly, the misfolded MHC molecules in activated T-cells have a signaling function \cite{Santos2004}. Additionally, the MHC molecule is known to misfold in several cases when in complex with TCR, and then it is is degraded by housekeeping enzymes \cite{Burr2011}. These articles show that the normally globular MHC molecules are known to misfold if in complex with the TCR molecule, containing the QKLVF subsequence. 

\subsection*{The GEYFT Prefix}
The GEYFT prefix (GLY,  GLU,  TYR,  PHE,  THR) appears 48 times in the NMR-identified members of the PDB\_Amyloid list, in the PDB entries 1OLG, 1SAE, 1SAF, 1SAK, 1SAL, 3SAK. These are non-amyloid structures. One of them, 1OLG is depicted in Figure 2. 

\begin{figure}[H]
	\begin{center}
		\includegraphics[width=12cm]{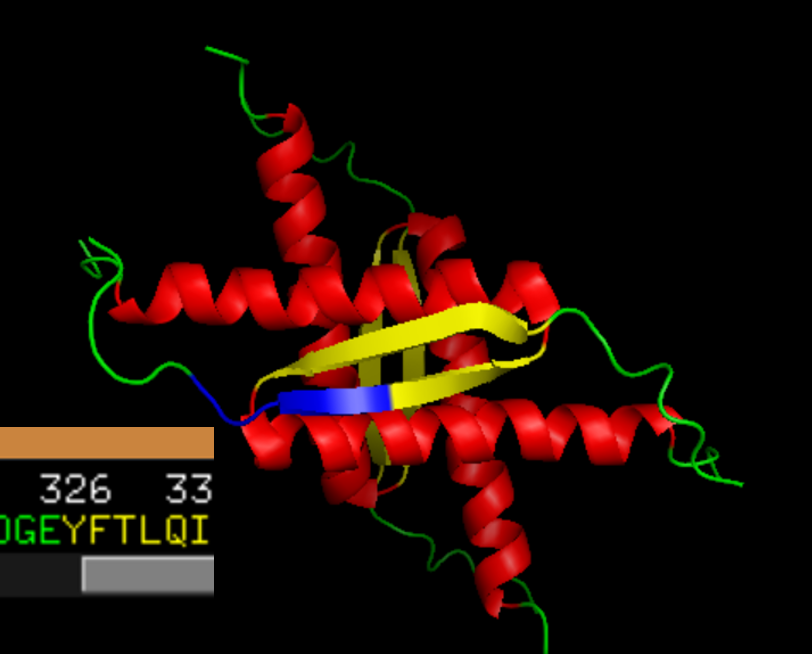}
		\caption{PDB entry 1OLG, depicted with PyMol. Yellow color denotes $\beta$-sheets. The GEYFT prefix, where GE are green (it is not a part of the parallel $\beta$-sheet) and YFT are yellow (i.e., YFT are the first three residues of the parallel $\beta$-segment), is emphasized at the left middle section of the figure, while its corresponding sequence at the lower left corner. }
	\end{center}
\end{figure}

Numerous GEYFT appearances in the PDB\_Amyloid list and also in the whole PDB are in p53 structures. p53 is a major tumor suppressor protein, whose gene is mutated in half of the human cancers  \cite{Vogelstein2000,Robles2010,Olivier2010}, and both its mutational deficiency in humans and the knock-out of its gene in mice imply early on-set cancers \cite{Malkin1990,Donehower1992}. It is very surprising that p53 mutations have a tendency of prion-like, contagious amyloid transitions: it is found that the amyloid-like aggregation plays a role in the loss of the p53 function in several organisms and cell types \cite{Sengupta2017,Silva2014,Silva2018}. 

We note that identifying non-amyloid p53 structures in the PDB\_Amyloid list shows the power of the methods by which the PDB\_Amyloid list was created  \cite{Takacs2019}: p53, an important non-amyloid structure with amyloidogenic properties is found in the list. We also note that numerous appearances of the GEYFT sequence in the whole PDB are also found in the p53 proteins.

Many GEYFT prefixes in the whole PDB are found in {\em Mycobacterium} (either {\em tuberculosis} or {\em smegmatis}) RNA polymerase transcription initialization complexes (e.g., 6DVC, 6JCX, 6JCY, 5ZX2). While it is not documented that these initialization complexes form amyloids, other bacterial transcriptional regulators do form amyloids. The {\em Bacillus subtilis} HeID, an RNA polymerase interacting helicase forms amyloids, as it was reported recently in \cite{Kaur2018}. Another finding that a mycobacterial global transcriptional factor, CarD, also forms amyloids, both {\em in vivo} and {\em in vitro} \cite{Kaur2018a}. Therefore, it would not be surprising if the GEYFT-containing mycobacterial RNA polymerase transcription initialization complexes also formed amyloid fibrils.

\subsection*{The HQKLV Prefix}
The HQKLV prefix (HIS, GLN, LYS, LEU, VAL) appears 25 times in the NMR-identified members of the PDB\_Amyloid list, in the PDB entries 2LMN, 2LMO, 2LMP, 2LMQ; these are all $\beta$-amyloid fibrils. If we search for the HQKLV subsequence in the whole PDB, we find numerous amyloid structures and some human amphiphysins: human amphiphysin isoform 1 (PDB codes 3SOG, 4ATM), and human BIN1/amphiphysin II (2FIC). Interestingly, the HQKLV subsequence appears in these pure $\alpha$-helix BIN1-structures as the part of the helix (Figure 3). 

\begin{figure}[H]
	\begin{center}
		\includegraphics[width=12cm]{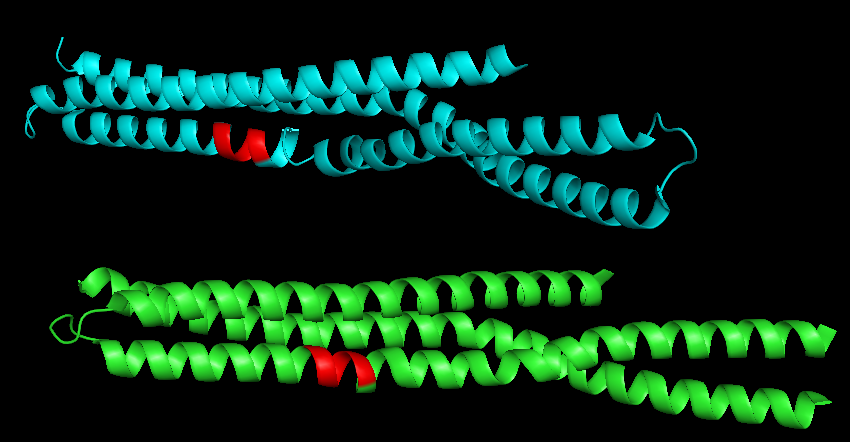}
		\caption{PDB entry 2FIC: the BAR domain of the human Bin1/amphiphysin II, depicted with PyMol. The red colored sections of the $\alpha$-helices correspond to the HQKLV subsequence. }
	\end{center}
\end{figure}		

BIN1 is not known to form amyloid-fibrils, but it is well-known to relate to late-onset Alzheimer's-disease: its gene is the second most important risk locus for Alzheimer's disease (after APOE: apolipoprotein E) \cite{Tan2013}, it is related to increased susceptibility for Alzheimer's disease \cite{Prokric2014}. More recently, it was shown that BIN1 regulates BACE trafficking and $\beta$-amyloid production. Therefore, we may conjecture that the HQKLV subsequence plays a role in amyloid-formation, even if it is in an $\alpha$-helix in BIN1 structures (Figure 3).

\subsection*{The GGERA Prefix}
The GGERA prefix (GLY. GLY. GLU, ARG, ALA) appears 106 times in the X-ray crystallography-identified members of the PDB\_Amyloid list, in the PDB entries 1DW9, 1DWK, 2IU7, 2IV1, 2IVQ, 4Y42; these are all bacterial cyanases. If we search for the GGERA subsequence in the whole PDB, the most interesting hit is the structure 1TCP: this is a tick anticoagulant peptide (TAP). The position of the GGERA subsequence is depicted in Figure 4.    

\begin{figure}[H]
	\begin{center}
		\includegraphics[width=12cm]{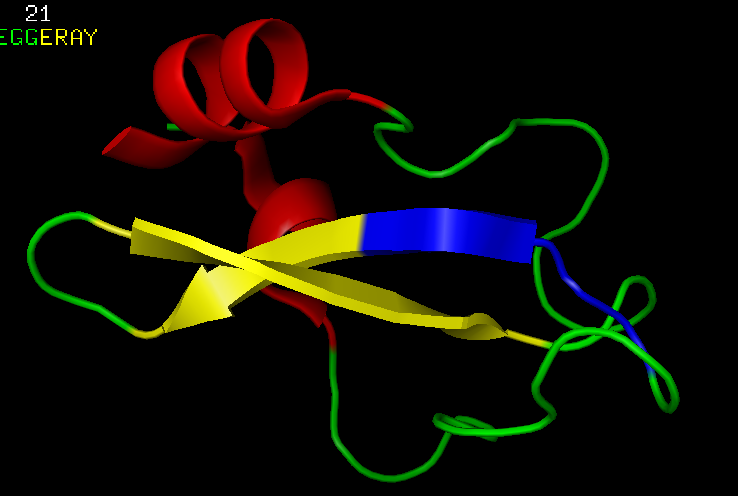}
		\caption{PDB entry 1TCP: the tick anticoagulant peptide (TAP), depicted with PyMol. The blue-colored section correspond to the GGERA prefix: GG preceeding the first three residues, ERA, in the $\beta$-sheet. }
	\end{center}
\end{figure}	

The $\beta$-sheet prefixes, listed above, were all related to prion- or amyloid-formation. Here, GGERA is found in an anti-coagulant molecule: the tick anti-coagulant peptide. Therefore, the GGERA subsequence is
\begin{itemize}
\item the prefix of the parallel $\beta$-sheet sections of several soluble proteins (cyanases) from the PDB\_Amyloid list, therefore the $\beta$-sheet, which starts with the GGERA sequence, is similar to those in the amyloid-structures, by satisfying properties (i), (ii) and (iii), listed in the Introduction;

\item but the cyanases 1DW9, 1DWK, 2IU7, 2IV1, 2IVQ, 4Y42 are all soluble proteins.
	\end{itemize}
Consequently, since GGERA also appears in the anti-coagulant 1TCP, it may have anti-amyloidogenic properties.

\section*{Conclusions}
 
By searching for the prefixes of the parallel $\beta$-sheet sections of the entries in the PDB\_Amyloid list in \url{http://pitgroup.org/amyloid/}, we were able to find numerous proteins in the whole PDB, from which only very recently were shown that they relate to the amyloid-formation. We conjecture that the prefixes listed may have structural roles in these amyloidogenic properties.

\section*{Data availability} 

The automatically updated PDB\_Amyloid web page is available at \url{http://pitgroup.org/amyloid/}. The list of the PDB codes of the PDB\_Amyloid can be viewed and downloaded at 
\url{http://pitgroup.org/apps/amyloid/amyloid_list}.

\section*{Acknowledgments}
 KT and VG were partially funded by the VEKOP-2.3.2-16-2017-00014 and the EFOP-3.6.3-VEKOP-16-2017-00002 programs, supported by the European Union and the State of Hungary. VG was partially funded by the NKFI-126472 and the NKFI-127909 grants of the National Research, Development and Innovation Office of Hungary. 
\bigskip 

\noindent Conflict of Interest: The authors declare no conflicts of interest.

\section*{Author contributions:} VG initiated the study, analyzed the results and wrote the paper. KT identified the parallel $\beta$-sheet segments in the spatial protein structures, and the prefixes in those $\beta$-sheets, satisfying the constraints, and described their appearances. 



\end{document}